# Strong coronal channelling and interplanetary evolution of a solar storm up to Earth and Mars

**Authors:** Christian Möstl[1,2]*, Tanja Rollett[1], Rudy A. Frahm[3], Ying D. Liu[4], David M. Long[5], Robin C. Colaninno[6], Martin A. Reiss[2], Manuela Temmer[2], Charles J. Farrugia[7], Arik Posner[8], Mateja Dumbović[9], Miho Janvier[10], Pascal Démoulin[11], Peter Boakes[2], Andy Devos[12], Emil Kraaikamp[12], Mona L. Mays[13,14], Bojan Vršnak[9]

**Affiliations:**

[1] Space Research Institute, Austrian Academy of Sciences, A-8042 Graz, Austria.

[2] IGAM-Kanzelhöhe Observatory, Institute of Physics, University of Graz, A-8010 Graz, Austria.

[3] Southwest Research Institute, 6220 Culebra Road, San Antonio, TX 78238, USA.

[4] State Key Laboratory of Space Weather, National Space Science Center, Chinese Academy of Sciences, Beijing 100190, China.

[5] Mullard Space Science Laboratory, University College London, Holmbury St. Mary, Dorking, Surrey, RH5 6NT, UK.

[6] Space Science Division, Naval Research Laboratory, Washington, District of Columbia, USA.

[7] Space Science Center and Department of Physics, University of New Hampshire, Durham, NH 03824, USA.

[8] NASA Headquarters, Washington, DC 20546, USA.

[9] Hvar Observatory, Faculty of Geodesy, University of Zagreb, 10 000, Zagreb, Croatia.

[10] Department of Mathematics, University of Dundee, Dundee DD1 4HN, Scotland, UK.

[11] Observatoire de Paris, LESIA, UMR 8109 (CNRS), F-92195 Meudon Principal Cedex, France.

[12] Royal Observatory of Belgium, 1180 Brussels, Belgium.

[13] Catholic University of America, Washington DC, USA.

[14] Heliophysics Science Division, NASA Goddard Space Flight Center, Greenbelt, Maryland, USA.

*Correspondence to: christian.moestl@oeaw.ac.at

**Abstract**

The severe geomagnetic effects of solar storms or coronal mass ejections (CMEs) are to a large degree determined by their propagation direction with respect to Earth. There is a lack of understanding of the processes that determine their non-radial propagation. Here, we present a synthesis of data from 7 different space missions of a fast CME which originated in an active region near disk center, and hence, a significant geomagnetic impact was forecasted. However, the CME is demonstrated to be channelled during eruption into a direction +37 ± 10° (longitude) away from its source region, leading only to minimal geomagnetic effects. In situ observations near Earth and Mars confirm the channelled CME motion, and are consistent with an ellipse shape of the CME-driven shock provided by the new Ellipse Evolution model. The results enhance our understanding of CME propagation and shape, which can help to improve space weather forecasts.

**Introduction**

Coronal mass ejections (CMEs) are the "hurricanes" of space weather and lead to massive disturbances in the solar wind plasma and magnetic fields through the inner heliosphere up to the heliospheric boundary[1,2,3]. During a CME, a mass of the order of $10^{12}$ kg is expelled from the Sun's corona, its outermost atmospheric layer, with kinetic energies of around $10^{23}$ J (ref. [1]). CMEs play a pivotal role in solar and space physics. They are responsible for the strongest disturbances in the geophysical environment, potentially leading to power blackouts and satellite failures at Earth[4]. Increasingly, policy makers recognize CMEs as a serious natural hazard, and counter-measures for the protection of space and ground-based assets are implemented.

A major requirement for producing reliable CME forecasts is to know their direction as accurately as possible as they propagate away from the Sun. Indeed, previous research[5-9] has found that various heliospheric structures may alter the CME trajectory to change its geomagnetic impact drastically. The strongest change in the propagation direction from its solar source position, which coincides with the flare for strong eruptions[10], has been mainly argued to occur within the first few solar radii where the magnetic forces acting on the CME are strongest[7,11,12,13], although possible changes of CME direction during interplanetary propagation have also been put forward[5]. Non-radial CME eruptions of up to 25° in longitude were predicted[8] due to the channelling of the CME to the location of the streamer belt, where the coronal magnetic field strength has a minimum. Coronal holes, which are regions of high magnetic field strength from where the fast solar wind emanates, are able to deflect CMEs away from their source regions, depending on their area and distance to the CME source[6,14,15]. However, precise quantifications and the maximum possible amount of deflection remain unclear. Deflected motions from high to low solar latitudes have been described for prominences, which upon eruption often form part of a CME[16,17].

A major obstacle for quantifying the CME deflection in heliospheric longitude is the lack of coronagraphs that can image CMEs from outside the ecliptic plane. Consequently, deflections in heliospheric longitude are much more difficult to analyze than those in latitude. To accurately quantify this process, it is thus necessary to study the complete chain of solar, coronagraphic, heliospheric and in situ observations. This is now possible with the multiple imaging and in situ spacecraft available, forming the Great Heliospheric Observatory.

In this work, we discuss an event that emphasizes the pressing need for improved real time predictions of the geomagnetic effects of CMEs. On 7 January 2014, a very fast CME erupted from a solar active region facing the Earth. Fast CMEs that erupt from source regions close to the solar disk center are usually expected to impact the Earth[10,18], so many observers around the world predicted that this CME would be strongly geo-effective and yet, no geomagnetic storm followed. We demonstrate that this CME was strongly channelled into a non-radial direction by the effects of its locally surrounding magnetic field rather than by coronal holes. We also show how the CME evolved in the heliosphere up to its arrival at Earth and Mars, confirming the inferences from solar imaging. To this end, we introduce the Ellipse Evolution (ElEvo) model for studying the CME propagation, which lets us derive constraints on the global shape of the CME-driven shock. To explain what happened in this event, we take a tour from the solar observations into interplanetary space, providing a synthesis of observations from 13 instruments on 7 space missions.

## Results

**Solar on-disk observations**

On 7 January 2014 19 UT, a CME erupted from a source region at 12° south and 8° west (S12W08) of disk center, accompanied by a plethora of phenomena such as a flare, coronal dimmings, a global coronal wave, and post-eruption arcades. These are the classical on-disk signatures of an erupting CME[19]. The flare, peaking at 18:32 UT, was of the highest class (X1.2). A very fast CME (projected speed of ~2400 km s$^{-1}$) was observed by the coronagraph instruments onboard the Solar and Heliospheric Observatory (SOHO)[20] in real time. Alerts for a G3 class geomagnetic storm or higher (on a scale from G1 to G5) were sent out by various space weather prediction centers around the globe[21] and were picked up by the media.

Figure 1a shows the state of the solar corona on 7 January 2014 19:30 UT, about 1 hour after the flare peak, imaged by the Solar Dynamics Observatory Atmospheric Imaging Assembly (SDO/AIA[22]) at 193 Å. Several active regions can be seen, with the largest one at the center of the solar disk. Post-eruption arcades, magnetic loops filled with hot plasma and counterpart signatures of an erupting CME magnetic flux rope, are visible at the flare site, located at the southwest corner of the large active region. Two large coronal holes (CHs) are visible in the northeastern quadrant. In Fig. 1b, the results from the automatic Coronal Pulse Identification and Tracking Algorithm (CorPITA[23]) visualize the location of the front of a global coronal wave, which is thought to be driven by the lateral expansion of the CME[24]. The algorithm could successfully track the wave almost exclusively in the southwest quadrant of the solar disk. This is confirmed by a visual identification of the wave in running difference movies (Supplementary Movie 1) showing an asymmetric wave propagation. A similar pattern can be seen in the coronal dimming regions in Fig. 1c (derived[25] from SDO/AIA 211 Å), which emphasize the evacuation of the corona at the locations of the footpoints of the erupting flux rope[19] (Supplementary Movie 2). The dimming appears earlier and is more cohesive in the southwest quadrant than in the northeast. Consequently, both the coronal wave and the dimming provide early suggestive evidence of the non-radial motion of the erupting CME to the southwest as seen from Earth. What could be the physical cause of this non-radial propagation?

**Influence of coronal holes and solar magnetic fields**

In Fig. 1a, the locations of two large CHs in the northeastern hemisphere of the disk seem consistent with the hypothesis that the CHs acted as to deflect the CME into the opposite direction. We took a closer look at this hypothesis within the framework of the so-called Coronal Hole Influence Parameter (CHIP[14,15]). The CHIP depends on the distance from the coronal hole center to the CME source region, the area of the coronal hole as well as on the average magnetic field inside the CH. It can be considered as a parameter describing how strongly the CME is pushed away from the coronal hole. The CHIP value for the CME event under study (see methods section) is $F = 0.9 \pm 0.2$ G, which is a factor of 3-5 below that necessary to deflect a CME originating close to the central meridian almost completely away from the Earth[14,15]. In particular, the distance of the CHs to the flare is comparably large[15], which lets us dismiss the hypothesis that the CHs are mainly responsible for deflecting this CME.

The explanation for the non-radial propagation may rather be found in the solar magnetic fields near the eruption. Figure 1d shows the line-of-sight component of the photospheric magnetic field (SDO / Helioseismic and Magnetic Imager (HMI)[26]). An active region (AR 11944) is right at disk center, a few degrees east of the flare position. The positive polarity (white) sunspot had a particularly strong vertical magnetic field of ~3000 G, which exceeds usual values[27] by a factor of about 1.5. The flare is located in between this strong sunspot and the small negative polarity at its southwest (Fig. 1d), and the erupting CME orientation (see next section) is consistent with the direction of the photospheric inversion line in between those two magnetic polarities. The negative polarity is almost surrounded by positive polarities so that a coronal magnetic null point and related separatrices are expected. The study of this magnetic topology is of primary importance to understand flare reconnection (e.g. flare ribbon locations). A further study, outside the scope of the present paper, including magnetic field extrapolation and data-driven MHD simulations, would be needed to understand precisely the role of the AR magnetic field complexity on the early CME development. However, since this topology is local, it is unlikely to be important for the CME development on scales larger than the AR, which is the focus of the present paper.

Next, looking at Figs. 1e-f a potential field source surface (PFSS[28]) model shows that the streamer belt of closed field lines is highly inclined with respect to the solar equator (typical of solar maximum), and runs from north to south right above the strong active region. The CME source region is not under the streamer, but close to an area of open flux further west (green field lines in Figs. 1e-f). This provides some evidence that the CME has erupted in the direction of least resistance in the solar global field[17], consistent with results of numerical simulations[12,13]. The solar observations thus imply that the strongly non-radial motion of this CME is due to a combination of 2 effects: (1) the strong nearby active region magnetic fields to the northeast, and (2) the open coronal field to the west of the source. Both processes acted to channel the CME to the southwest of the solar disk, which was reflected in the asymmetries of the global coronal wave and dimmings.

**Coronal evolution**

We now take a look at multi-viewpoint coronagraph observations of the CME in Fig. 2a. We used two methods to estimate the CME propagation direction up to 30 solar radii ($R_\odot$). The first is the Graduated Cylindrical Shell (GCS) model, by which a wire-grid of a tapered hollow tube is fitted onto coronagraph images[29,30] by manual variation of several parameters controlling its shape. The triple viewpoints from the STEREO-B (COR1/2[31]), SOHO (C2/C3) and STEREO-A (COR1/2) imagers constrain the results very well[30]. At the time of the event, the two STEREO spacecraft were 151° ahead (A) and 153° behind (B) in heliospheric longitude with respect to Earth, at distances of 0.96 AU (A) and 1.08 AU (B). The CME propagates to the east in STEREO-B, where the event is seen as backsided, which shows that the CME longitude must be greater than -153° +180° = 27° west of the Sun-Earth line. The GCS model was applied between 7 January 18:15 and 19:30 UT, when the resulting model apex position was between 2.1 and 18.5 $R_\odot$. The average 3D speed of the CME apex was 2,565 ± 250 km s$^{-1}$, derived from a linear fit to $R(t)$, not far from the fastest speeds ever observed[3] (~ 3,000 km s$^{-1}$). A constant CME direction is consistent with the time evolution in the coronagraph images, which gives 32 ± 10° (west) and -25 ± 5° (south, with quoted errors common for the method[30,32]). This means that already very close to the Sun, at 2.1 $R_\odot$, the final direction of the CME was attained.

A 2$^{nd}$ method was used to find the speed of the CME segment that propagates in the ecliptic plane. We applied a triangulation technique[33] to the CME leading edge in SOHO/C2/C3 and STEREO-A/COR2/HI1[34] observations. Our results are averages of 2 methods (Fixed-$\beta$ and Harmonic Mean) assuming a small and wide CME extent in the ecliptic along the line of sight[33]. From a linear fit to $R(t)$ between 20-30 $R_\odot$, we find a speed of 2,124 ± 283 km s$^{-1}$, slightly lower than the apex speed from the GCS method. The direction of the CME front above 20 $R_\odot$ is W45 ± 10°. This is further west from Earth by about 13° compared to the GCS results, which can be expected since parts of CMEs seen at different latitudes may travel in slightly different directions, because of the CME 3D tube shape. This is reasonable, as Fig. 2a shows the CME to be oriented with a moderate inclination angle to the ecliptic. Further from the Sun, we tracked the CME ecliptic leading edge to about 25° elongation with the STEREO-A Heliospheric Imager (HI), and applied the Fixed-Phi-Fitting method[32]. This results in a speed of 2,131 ± 210 km s$^{-1}$, consistent with triangulation. We use this as the initial speed for further modeling of the shock evolution in the ecliptic plane. We also assume that the CME leading edge is representative of the position of the CME-driven shock, which has been confirmed with imaging of CMEs at large elongations from the Sun and their in situ observations[35]. For the CME direction, we take 45 ± 10° west of the Earth or 37 ± 10° away from the source region in heliospheric longitude, exceeding expected values for non-radial CME eruption in the corona[8].

**Interplanetary evolution**

Fig. 2b extends the $R(t)$ and $V(t)$ functions of the CME shock up to Mars (1.66 AU). The interplanetary kinematics towards Earth and Mars are shown together with the arrival times at both planets, which will be discussed further below. We modeled the shock kinematics with the drag-based-model (DBM[36]), which analytically describes the deceleration of CMEs by using equations of aerodynamic drag. It gives similar performances for arrival time predictions of CMEs at Earth as numerical simulations[37]. The DBM has 2 free, constant parameters: the drag parameter $\gamma$ and the background solar wind speed $w$. Parameter $\gamma$ contains information on the CME mass and size, the ambient solar wind density and the interaction between the CME and the solar wind[36]. For the background solar wind we use $w$ = 400 km s$^{-1}$, an average solar wind speed observed at Earth by the Wind spacecraft a few days around the time of the CME. This means that there are no high-speed solar wind streams west of Earth near the CME principal direction. A previous CME on 6 January 2014 was directed towards STEREO-A[38], and is not expected to influence the propagation of the 7 January CME towards Earth and Mars. Both inferences support the view that we can safely use a constant direction of motion and constant $\gamma$ and $w$. Parameter $\gamma$ may vary[36] from 0.1 to $2 \times 10^{-7}$ km$^{-1}$. Because Mars is close to the apex of the ecliptic part of the CME shock (6° away), we can choose a value for $\gamma$ that makes $R(t)$ consistent with the shock arrival time at Mars ($t_{Mars}$, see next section). This results in $\gamma = 0.165 \pm 0.005 \times 10^{-7}$ km$^{-1}$, an expected value[36] for a fast CME propagating into a slow, unstructured solar wind. The time $t_{Mars}$ has an uncertainty of ± 1 hour (see below), providing the error margin in $\gamma$. We now constrain the interplanetary shock evolution further with a new model for the shape of the CME shock in combination with the available in situ observations.

Multipoint in situ observations of interplanetary CME (ICME) signatures, at longitude differences[39,40] from 10° to the size of the CME shock of ~100° can give constraints on how a CME evolves in the interplanetary medium[41-46]. Few such events have been described in the literature, and thus the global shape and extension of CMEs are poorly known. In a fortunate coincidence on 7 January 2014, Mars was at a heliospheric longitude of 51° west of the Earth. Judging from the CME principal direction, Mars and Earth should see the apex and flank of the CME, respectively. Fig. 3a visualizes the position of the planets and the STEREO spacecraft in the ecliptic plane, together with the shock evolution up to its arrival at Mars (see Supplementary Movie 3). We use the new Ellipse Evolution (ElEvo) model for describing the global shape of the interplanetary shock (see methods section) which is based on statistics of single-point shock observations[47]. Before we discuss the range of possible parameters and errors for this model, we take a look at the solar wind and planetary in situ data, serving as boundary conditions for ElEvo.

**Arrival at Earth**

Figure 3b-g shows the solar wind magnetic field and bulk plasma parameters observed by the Wind spacecraft at the L1 point near Earth. For a very fast CME from a source region near disk center, such as the 7 January 2014 event, the expected in situ ICME signatures are a shock, followed by a sheath region of enhanced plasma density and temperature, and a magnetic flux rope with a size of the order of 0.1 AU in the radial direction[48]. However, in situ observations are limited to those acquired along a spacecraft trajectory through the global CME structure, and for a glancing encounter, the flux rope is likely to miss the spacecraft[49]. On 9 January 2014 at 19:40 UT, an interplanetary shock hit Wind, causing a sudden jump in the solar wind speed from 390 km s$^{-1}$ to 465 km s$^{-1}$. This time, which we label as $t_{Earth}$, defines the CME shock arrival at Earth. No other CME can explain the arrival of this shock. Fig. 2b demonstrates that the arrival speed of the shock given by ElEvo at Earth indeed matches the in situ shock speed of 488 km s$^{-1}$, derived from the magnetohydrodynamic Rankine-Hugoniot relations. The shock was weak, with a magnetosonic Mach number of 1.2. The orientation of the shock using the co-planarity theorem

points in the radial direction away from the Sun and 26° northward, which is consistent with the main direction of the CME to the south of the ecliptic plane[45]. A sheath region followed, with an average speed of 417 ± 20 km s$^{-1}$ and elevated proton temperatures, extending up to 10 January 2014 06:00 UT. This region has a radial size of 0.104 AU, with a magnetic field of 9.5 ± 1.9 nT, which is enhanced compared to average solar wind values[48]. This magnetic field is relatively weak compared to the average field usually found in ICMEs which contain flux ropes[48].

In summary, the above indicates that this CME almost entirely missed the Earth, because a shock-sheath pair is seen but not any type of magnetic ejecta. Fig. 3g shows the corresponding Disturbance storm time (Dst) index, which is derived from a combination of equatorial ground station magnetometers around the world. The sheath region has a magnetic field and radial size which is comparable to non-cloud ejecta[48] at 1 AU, but due to the predominantly northward $B_z$ (Fig. 3c), Dst does not even reach levels typical of a minor geomagnetic storm (Dst < -50 nT). The maximum of the *Kp* index was only 3, which is below the NOAA threshold for a G1 category geomagnetic storm. However, the CME lead to a major solar energetic particle event near Earth, which resulted in an S3 solar radiation storm on the NOAA scale from S1-S5.

It is also interesting to note that the particular orientation of the CME favors a miss of the flux rope at Earth too, because the east "leg" is far below the ecliptic, as demonstrated by the SOHO image in Fig. 2a. Thus, in addition to the non-radial eruption, this orientation should have contributed to the false forecasts, because CMEs with a moderate to high inclination of their axes to the ecliptic plane have a small angular extent in the ecliptic[39], making it likely that the flux rope inside this CME has crossed the ecliptic to the west of the Earth. As a consequence, the possibly strong southward magnetic fields of the CME flux rope have not impacted the Earth's magnetosphere at all.

**Arrival at Mars**

Fig. 4a shows an electron spectrogram by the Mars Express (MEX) Electron Spectrometer (ELS) instrument[50]. Enhanced fluxes of electrons, originating from both the planet's ionosphere and the solar wind, are indicated by red colors in Fig. 4a, are seen starting on late 10 January, and they show a drastic change on 11 January, when the electrons became both more intense and reached higher energies. These data suggest that a CME is arriving on this day at Mars, because the CME sheath contains both denser and faster plasma than the surrounding slow solar wind and additional energy is added into the induced magnetosphere of Mars by the CME. The MEX data do not allow a more precise definition of the arrival time than 11 January ± 1 day, so we turn to data from the Mars surface by the Radiation experiment (RAD[51]) onboard the Mars Science

Laboratory's Curiosity rover. RAD is able to observe Forbush decreases (FDs), a temporary decline in galactic cosmic ray intensity when an ICME passes a detector[52,53]. Figure 4b shows RAD count rates per second of energetic particles, which includes primary particles and secondary ionizing radiation created by solar and galactic cosmic rays in the Mars atmosphere and regolith. After a solar energetic particle event, related to a different CME[38], a return to normal counts is seen followed by a Forbush decrease onset at $t_{FD}$ = 10 January 2014 22:30 UT ± 1 hour. What appears to be a single-step behavior of the FD in Fig. 4b could indicate an arrival of only the ICME shock, and not the flux rope[53]. How is the time $t_{FD}$ related to the ICME shock arrival at Mars? At Earth, there is in general a relatively tight correspondence in timing[54] between ICME arrivals at L1 and ground-based FD onsets, with the shock mostly arriving a few hours before the FD onset. Given that the corresponding FD onset at Earth (not shown) is within < 2 hours of the shock arrival at the Wind spacecraft, the start of the FD at Mars can be reasonably assumed to coincide, within a 2-hour window, with the arrival at Mars of the presumably strong shock driven by this fast CME: $t_{Mars} = t_{FD}$.

**Global shape of the shock**

As we now know the arrival times of the CME at Earth and Mars, we can go back to Fig. 3a which shows the shape for the CME shock in the ecliptic plane given by ElEvo for equidistant timesteps. The model assumes that the CME shock propagates along a constant direction with a constant ellipse aspect ratio ($a_r$), constant angular half width ($\lambda$), and one main axis of the ellipse oriented along the radial direction from the Sun. It is also possible to calculate the speeds and arrival times at in situ locations along the ellipse front analytically (see Methods section). The evolution of the ellipse apex, which is the point of the ellipse farthest from the Sun along the CME central direction, is modeled with the DBM. From an optimization analysis (see Methods section) using the multipoint in situ arrival times we find the ellipse aspect ratio to vary as $a_r$ =1.4 ± 0.4, for half widths of the ellipse ranging from 35 to 60° (under the condition that the ellipse always hits Earth) and for the central ellipse direction at W45 ± 10°. Plotted in Fig. 3a is an ellipse with parameters $a_r$ =1.43, $\lambda$ = 50° and direction W45. The multipoint modeling of the CME shock thus implies an elliptical shape elongated perpendicular to the propagation direction[47].

## Discussion

We presented for the first time the full evidence for a very strong non-radial motion of a coronal mass ejection, with complete observations from the Sun and two planetary impacts. The CME from 7 January 2014 19:00 UT almost entirely missed Earth despite its source being close to the center of the solar disk. We attributed this to a non-radial CME propagation direction, which was attained very close to the Sun (< 2.1 $R_\odot$), rather than to a deflection in interplanetary space[5]. The observations do not show a "deflection", which implies a change in direction, but rather a "channelled" CME motion, which is non-radial starting already with its inception on the Sun. We found a surprisingly large magnitude of this channelling with respect to the source region on the Sun (+37 ± 10° in heliospheric longitude), and a so far largely unrecognized process causing the channelling by nearby active region magnetic fields[49] rather than coronal holes[6]. The observations emphasize the need to understand the interplay between the active region and global magnetic fields in order to better predict the direction of CMEs, and support previous studies which derive altered CME trajectories from modeling the background coronal magnetic field[7,8,11].

We also showed suggestive evidence that the non-radial CME motion can be seen in extreme ultraviolet (EUV) observations of the corona, which showed asymmetries in the global coronal waves and dimming regions with respect to the flare position. Because such asymmetries can also arise from the structure of the solar corona[55], further research is needed on the possibility of diagnosing non-radial CME motions within EUV images. Finally, the arrivals of the CME-driven shock have been observed in situ by the Wind spacecraft near Earth and by the RAD experiment on the Mars Science Laboratory's Curiosity rover on the Mars surface. These observations, together with results provided by the new Ellipse Evolution (ElEvo) model, show that these arrivals are consistent with the CME direction given by solar and coronagraph imaging. The in situ arrival times allowed to directly constrain the global shape of the CME-driven shock to an ellipse with aspect ratio of 1.4 ± 0.4, with the ellipse elongated perpendicular to the direction of CME motion[47].

The enhanced understanding of non-radial CME propagation presented in our paper will be helpful for real time CME forecasting[18] in order to avoid false positive predictions, as it was the case for the event studied, i.e. a CME that was predicted to impact the Earth actually missed it. However, this means that a false negative forecast is also possible: a CME that is launched from a source region as much as 40° away in longitude from the Sun-Earth line may impact Earth centrally. It needs to be emphasized that non-radial CME motion in heliospheric longitude is very difficult to study because the images showing the CME radial distance from the Sun result from integrations along the line of sight. The upcoming Solar Orbiter mission[56], imaging for the first time the Sun and the heliosphere from outside of the ecliptic plane, will provide better insights into the dynamical processes responsible for non-radial CME eruptions.

In summary, the presented observations demonstrate the high value of many different instruments in spatially distant locations to study solar storms. We were able to draw a consistent picture of the evolution of a CME from its inception on the Sun, during which the event under study experienced a strongly non-radial motion, to the impacts at planets and spacecraft. These fundamental results should help to improve the reliability of real-time forecasts of space weather.

## Methods

**Calculating the coronal hole influence parameter**

The Coronal Hole Influence Parameter (CHIP) is given as[14]

$$\mathbf{F} = \left(\frac{\langle B \rangle A}{d^2}\right) \mathbf{e}_\mathbf{F}, \qquad (1)$$

with $\langle B \rangle$ the average magnetic field inside the coronal hole (CH), $A$ the CH area corrected for projection, $d$ the distance from the CH center to the eruption site and $\mathbf{e}_\mathbf{F}$ the unit vector along this direction. This vector defines the direction in which an erupting CME will be deflected due to the presence of the coronal hole, under the condition that $\mathbf{F}$ is sufficiently large. The values for $A$ and the barycenter position (the center of the CH region weighted by pixel intensity) were provided by the Heliophysics Event Knowledge base (HEK[57]), obtained with the SPoCA algorithm[58] used on SDO/AIA 193 Å images[22]. We calculated the distance from the CH barycenter to the eruption site (S12W08) with the great circle distance on the solar sphere. The areas of the northern (southern) CHs are $1.2 \pm 0.04 \times 10^{11}$ km$^2$ ($0.3 \pm 0.01 \times 10^{11}$ km$^2$), the distances to the source are $7.0 \times 10^5$ km ($4.5 \times 10^5$ km), and the average magnetic fields are 1.38 (3.76) G (measured from SDO/HMI[26]). As there are 2 coronal holes present, the CHIP is treated as a vector, which is summed up for all CHs present on the disk[14].

The resulting CHIP from both coronal holes at the location of the source region at S12W08 is $F = 0.9 \pm 0.2$ G, with the uncertainties arising from the calculation of the area with different thresholding techniques and an uncertainty in the determination of the magnetic field. The direction $\mathbf{e_F}$ arising from the CHIP analysis is towards a position angle of about 230° (PA is measured from solar north at 0° to east – left side in a solar image – at 90°, south at 180° and west at 270°), which also seems at first glance consistent with the CME propagation direction to the southwest, but the low CHIP[14,15] means that the combined coronal holes are not sufficiently close, have a large enough area or strong magnetic field to explain the non-radial propagation of the CME.

**Derivation of the ElEvo model for the evolution of CMEs**

To model the shape of the shock driven by the CME in the interplanetary medium, we introduce a new method which describes the shock as an ellipse in the ecliptic plane, which we call the Ellipse Evolution (ElEvo) model. It allows the extension of 1D models for heliospheric CME propagation[36,59], which provide the distance-time $R(t)$ and speed-time $V(t)$ functions of the CME front, into 2D models of the evolution of CME boundaries in the ecliptic plane with only a few lines of code. ElEvo can thus be used to visualize the shape of a CME shock between the Sun to a given planet or in situ observing spacecraft using fully analytical formulas. Further, the speed and arrival time of any point along the ellipse can be calculated analytically.

ElEvo is an extension to a model describing CME boundaries as self-similar expanding (SSE) circles[60,61]. Whereas SSE has been designed to derive CME parameters from observations of CMEs at large angles from the Sun with a heliospheric imager[32,61], it can also be used to propagate a CME into the heliosphere and calculate expected planetary arrivals and speeds[60]. The initial speed, direction and width of a CME, which are known from coronagraph observations, can be used as initial conditions. The advantage of ElEvo is that the shape of an ellipse is more flexible than a SSE circle, and thus, better suited for consistent modeling with multipoint in situ observations because the aspect ratio of the ellipse is a free parameter.

The assumptions of the ElEvo model are: (1) the angular width in heliospheric longitude of the CME boundary remains the same for all times, (2) one principal axis of the ellipse is oriented along to the propagation direction, and (3) the ellipse aspect ratio and (4) direction are both constants. However, it is possible to change the aspect ratio and direction in a code as a function of time, but this is not implemented in the current study. For describing the interplanetary deceleration of the CME, we use the drag based model (DBM[36]) which describes the kinematics for the distance $R(t)$ and the speed $V(t)$ of the CME as function of time ($t$). It has two free parameters: the drag parameter $\gamma$ (on the order of 0.1 to $2 \times 10^{-7}$ km$^{-1}$) which describes the amount of drag exerted by the solar wind on – strictly speaking – the CME flux rope, and $w$ which is an average of the background solar wind speed. By choosing low values of $\gamma$ it is possible to describe the shock propagation up to 1 AU with DBM, so it can be used to calculate CME shock arrival times and speeds. In summary, the ElEvo model in combination with the DBM has 4 free parameters: (1) the inverse aspect ratio $f$, (2) the ellipse half angular width $\lambda$, (3) the drag parameter $\gamma$ and (4) the background solar wind speed $w$. We now derive the equations necessary to code the geometry of the ElEvo model to visualize the self-similar expanding ellipse as it propagates away from the Sun as well as the speed for each point along the ellipse front.

**Visualizing a self-similar propagating ellipse:** Figure 5a shows the geometry of an ellipse under the assumptions described above. The $R(t)$ of the ellipse apex (the point of the ellipse farthest from the Sun along the ellipse central direction) is given by DBM[36]. In this section, we derive equations for the ellipse semi-major axis $a$ and semi-minor axis $b$ as a function of $R(t)$, the inverse aspect ratio $f$ and the half width $\lambda$. We use $f = b / a$ rather than $a_r = a / b$ because it simplifies the following calculation. The equations

$$\begin{aligned} f &= b/a, \\ \beta &= \lambda, \\ \theta &= \arctan(b^2 / a^2 \tan \beta), \end{aligned} \qquad (2)$$

follow from the definition of $f$ and the definition of angle $\beta$, which is the angle between the semi-major axis $a$ and the normal to the tangent at point $T$ (Fig. 5a). The location of $T$ is the point of tangency on the ellipse for a line originating at the Sun. It can be easily seen that $\beta = \lambda$ by checking the sum of the angles of the small orange triangle in Fig. 5a in relation to a larger triangle (not highlighted) containing the angles $\lambda$ and $\eta$. The polar angle of the ellipse $\theta$ is given by a relationship from general ellipse geometry between $\beta$ and $\theta$. It is important to emphasize that we further construct the ellipse based on this particular value of $\theta$, for which a line with distance $r$ connects the ellipse center $C$ to point $T$. Combining equations (2) gives a relationship for the polar angle $\theta$ based on known parameters,

$$\theta = \arctan(f^2 \tan \lambda). \qquad (3)$$

From the law of sines on the large orange shaded triangle in Fig. 5a we derive:

$$\frac{\sin \lambda}{r} = \frac{\sin \alpha}{R(t) - b}. \qquad (4)$$

Angle $\alpha$ follows from the angles of the orange shaded triangle, and distance $r$ from the definition of an ellipse in polar coordinates:

$$\alpha = 90° + \theta - \lambda,$$
$$r = ab / \sqrt{(b\cos\theta)^2 + (a\sin\theta)^2}.$$
(5)

The last equation can be rewritten with the definition of $f$ as

$$\omega = \sqrt{(f^2 - 1)\cos^2\theta + 1},$$
$$r = b/\omega.$$
(6)

Introducing $\alpha$ from equation (5) and the last equation for $r$ into equation (4) then eliminates the unknowns ($\alpha$, $r$) and expresses $b$ in function of known variables:

$$b = \frac{R(t)\omega\sin\lambda}{\cos(\lambda - \theta) + \omega\sin\lambda},$$
$$a = b/f,$$
$$c = R(t) - b.$$
(7)

Equations (7) are the final description of the ellipse parameters. The minor axis $b$ of the ellipse depends on all known variables ($R(t), f, \lambda$) through $\theta$ and $\omega$, from equations (3) and (6). The major axis $a$ then simply follows from the definition of $f$ in equation (2). The heliocentric distance of the center of the ellipse is parameter $c$ (Fig. 5b), closing the model equations necessary for visualizing the ellipse.

**Calculation of speeds along the ellipse front:** For comparison to in situ observations, which give parameters such as the speed and arrival time of the CME shock with very good accuracy[48], one needs to know the speed of any point along the ellipse front as a function of the ellipse parameters. This problem has been solved analytically for the circular SSE geometry[60], and we introduce here the corresponding analytic solution for ellipses.

Figure 5b demonstrates the geometry, with $\Delta$ being the known angle between the CME central direction and for example the Earth, which could also be any other planet or spacecraft in the solar wind. The direction of the apex with respect to a coordinate system including the Sun and Earth depends on different methods for CMEs observed with coronagraphs[29,33] or heliospheric imagers[61,62].

We introduce a coordinate system centered on the ellipse (Fig. 5b), with coordinate $X$ being perpendicular to the CME propagation direction and $Y$ orthogonal to $X$. Here, **c** is the vector from the Sun to the ellipse center, **d** connects the Sun to the front edge of the ellipse in the direction of Earth (i.e., **d** stops at the ellipse boundary and does not connect the Sun to the planet), and **r** connects the center to the end point of **d** on the ellipse:

$$\begin{aligned}\mathbf{c} &= (0,c), \\ \mathbf{d} &= \mathbf{c} + \mathbf{r}.\end{aligned} \quad (8)$$

The problem consists in finding the norm of **d** as a function of $\Delta$. There are 2 crossings of **d** with the ellipse, one at the rear and one at the front (Fig. 5b), which will form the two solutions of the problem. The coordinates of **r** can be expressed with the projections of the vector **d-c** in the $X/Y$ coordinates:

$$\begin{aligned}\mathbf{r} &= \mathbf{d} - \mathbf{c}, \\ d_x &= d \sin(\Delta), \\ d_y &= d \cos(\Delta), \\ r_x &= d_x, \\ r_y &= d_y - c.\end{aligned} \quad (9)$$

Then, $r_x$ and $r_y$ can be introduced into the definition of an ellipse in cartesian coordinates,

$$\left(\frac{r_x}{a}\right)^2 + \left(\frac{r_y}{b}\right)^2 = 1. \quad (10)$$

This results in the following expression, which was simplified with the definition of $f$ from equation (2):

$$d^2 f^2 \sin^2 \Delta + (d \cos \Delta - c)^2 = b^2. \tag{11}$$

This quadratic equation gives two analytic solutions of the front and rear crossings of $d$ with the ellipse as a function of the parameters $b$, $c$ from equation (7), $f$ from equation (2) and $\Delta$:

$$d_{1,2} = \frac{c \cos \Delta \pm \sqrt{(b^2 - c^2) f^2 \sin^2 \Delta + b^2 \cos^2 \Delta}}{f^2 \sin^2 \Delta + \cos^2 \Delta}. \tag{12}$$

The solution with the positive sign in front of the root is the "front" solution ($d_1$) and the one with the negative sign the "rear" solution ($d_2$). The speed $V_\Delta(t)$ of the ellipse at the position defined by the angle $\Delta$ is derived from the self-similar expansion of the ellipse, which implies a constant half width $\lambda$. The assumption of self-similar expansion means that the shape of the ejection must not change in time, so the ratio between speeds and distances for all points along the ellipse must be constant[60]:

$$V_\Delta(t) = \frac{d_1(t)}{R(t)} V(t). \tag{13}$$

Further, the time when $d_1(t)$ is equal to the heliocentric distance of the planet or in situ observer gives the arrival time of the ellipse at the in situ location, and the speed $V_\Delta(t)$ at this time the arrival speed.

**Analysis of the aspect ratio and width of CME shocks**

In this section we discuss how to find optimal solutions for the ElEvo shape when multipoint observations of the ICME arrival are available. We first create a shock apex kinematic $R(t)$ with the DBM with parameters $\gamma = 0.165 \times 10^{-7}$ km$^{-1}$, $w = 400$ km s$^{-1}$ which yields a DBM arrival time at Mars consistent with the observed arrival time $t_{\text{Mars}}$. From the $R(t)$ apex, we calculated with equation (12) for a range of half widths from $45° < \lambda < 60°$ the parameter $d_{1,\text{Earth}}$ for the longitude of Earth, at the Earth arrival time $t_{\text{Earth}}$. This range for values of $\lambda$ is chosen because the half shock extension is thought to be around 50° in heliospheric longitude[40]. From the in situ observations we know that the shock has impacted the Earth. Thus, for an ellipse apex direction of W45 ± 10°, $\lambda$ must be larger than 45° ± 10°, or the shock would not reach Earth. This also means that the half width ($\lambda$) of the shock is > 35° in the ecliptic, consistent with the value of ~50°.

Because the ellipse shape needs to be consistent with both the Earth and Mars arrival times, we repeated the same procedure for Mars and calculated the average of the residual distances $D$ between the ellipse front ($d_1$ for the corresponding values of $\Delta$ for Earth and Mars) and the heliocentric distances of both planets ($d_{Earth}$ and $d_{Mars}$) at the respective observed arrival times $t_{Earth}$ and $t_{Mars}$:

$$\Delta d_{Earth} = |d_{Earth} - d_{1,Earth}|, \qquad (14)$$
$$\Delta d_{Mars} = |d_{Mars} - d_{1,Mars}|,$$
$$D = (\Delta d_{Earth} + \Delta d_{Mars})/2.$$

Figure 6 shows parameter $D$ as function of $a_r$ varied from 0.5 to 2.5 and for different $\lambda$ indicated by colors. This plot was made with the average CME direction of W45. For each $\lambda$, an optimal solution for $a_r$ exists where $D$ has a minimum. At the minimum, the ellipse impacts the Earth and Mars at the observed arrival times. For two in situ spacecraft that observe an ICME arrival, there are optimal solutions for pairs of $\lambda$ and $a_r$ that match the 2 given in situ arrival times for the CME shock if either $\lambda$ or $a_r$ is known or assumed. We note in passing that for CME events where 3 or more in situ arrival times would be available, there would be higher residuals, and a stronger constraint might be derived. The important new result is that $\lambda$ and $a_r$ are not independent of each other, and the result of the optimization procedure is a range of $1.69 > a_r > 1.23$ for half widths of $45° < \lambda < 60°$ when keeping the direction constant at W45. For a half width of 50°, the optimal $a_r$ =1.43 (used in Fig. 3a). In general, as the width increases the aspect ratio must become smaller to be consistent with the in situ arrival times.

To fully include the errors from the direction determination we also experimented with varying the shock apex direction by ± 10° around W45, a typical error for CME directions by the triangulation method[33]. For the W55 case, $55° < \lambda < 60°$, the optimal aspect ratio is in the range $1.84 > a_r > 1.58$. For the W35 case, $35° < \lambda < 60°$, the aspect ratio is $1.51 > a_r > 0.99$. Thus, including the errors from both the direction and varying the half width within reasonable values leads to a considerable possible variation in $< a_r > = 1.4 \pm 0.4$. However, $< a_r >$ indicates the global shock shape to be a slightly elongated ellipse, being only close to a circular shape for a very extreme choice of parameters.

To better visualize this optimization process, we illustrated in Fig. 3a a distance window named "shape constraint" along the Sun-Earth line where the ellipse at the Mars arrival time has to pass through in order to be consistent with the shock arrival time at Earth. After impacting the Wind spacecraft, the shock traveled 26 h 50 min ± 1 h until its apex hit Mars. We assume that during this time the shock traveled with a speed of 488 km s$^{-1}$, which is close to the slow solar wind speed so we do not expect much deceleration. With such a speed, the shock was at $t_{Mars}$ at a distance of 0.315 ± 0.015 AU further away from Earth, along the Sun-Earth line. This distance has an error (indicated on the figure by small horizontal lines on Fig. 3a) due to the uncertainty of ± 1 hour in $t_{Mars}$. One can see in Fig. 3a that the outermost (red) ellipse indeed crosses the "shape constraint" window, which means that the implementation of ElEvo is consistent with the observed multipoint arrival times.

## Acknowledgments

This study was supported by the Austrian Science Fund (FWF): [P26174-N27, V195-N16]. T. R. gratefully acknowledges the JungforscherInnenfonds of the Council of the University Graz. D.M.L is a Leverhulme Early-Career Fellow funded by the Leverhulme Trust. M.D. and B.V. acknowledge financial support by the Croatian Science Foundation under the project 6212 SOLSTEL. Y.D.L. was supported by the Recruitment Program of Global Experts of China, NSFC under grant 41374173 and the Specialized Research Fund for State Key Laboratories of China. The presented work has received funding from the European Union Seventh Framework Programme (FP7/2007-2013) under grant agreement No. 606692 [HELCATS] and No. 284461 [eHEROES]. Part of this work was supported by NASA grants NNX13AP39G, NNX10AQ29G and STEREO Farside Grant to UNH. MEX/ASPERA-3 is supported in the USA by NASA contract NASW-00003. RAD is supported by the National Aeronautics and Space Administration (NASA, HEOMD) under Jet Propulsion Laboratory (JPL) subcontract #1273039 to Southwest Research Institute and in Germany by DLR and DLR's Space Administration grant numbers 50QM0501 and 50QM1201 to the Christian Albrechts University, Kiel. A. D. acknowledges support from the Belgian Federal Science Policy Office through the ESA-PRODEX program, grant No. 4000103240. E.K. acknowledges support from the European Commission's Seventh Framework Programme (FP7/2007- 2014) under the grant agreement nr. 263506 (AFFECTS project), and grant agreement nr. 263252 (COMESEP project). This research has made use of the Heliophysics Event Knowledge (HEK) database and the ESA JHelioviewer software. We thank Janet G. Luhmann and Julia K. Thalmann for discussions, and the center for geomagnetism in Kyoto for providing the Dst indices.

## Author contributions

C. M. has designed and coordinated the study. All authors have contributed to the analysis of data, the creation of the figures and to writing the manuscript.

We declare that none of the authors has any competing financial interests.

**Figures**

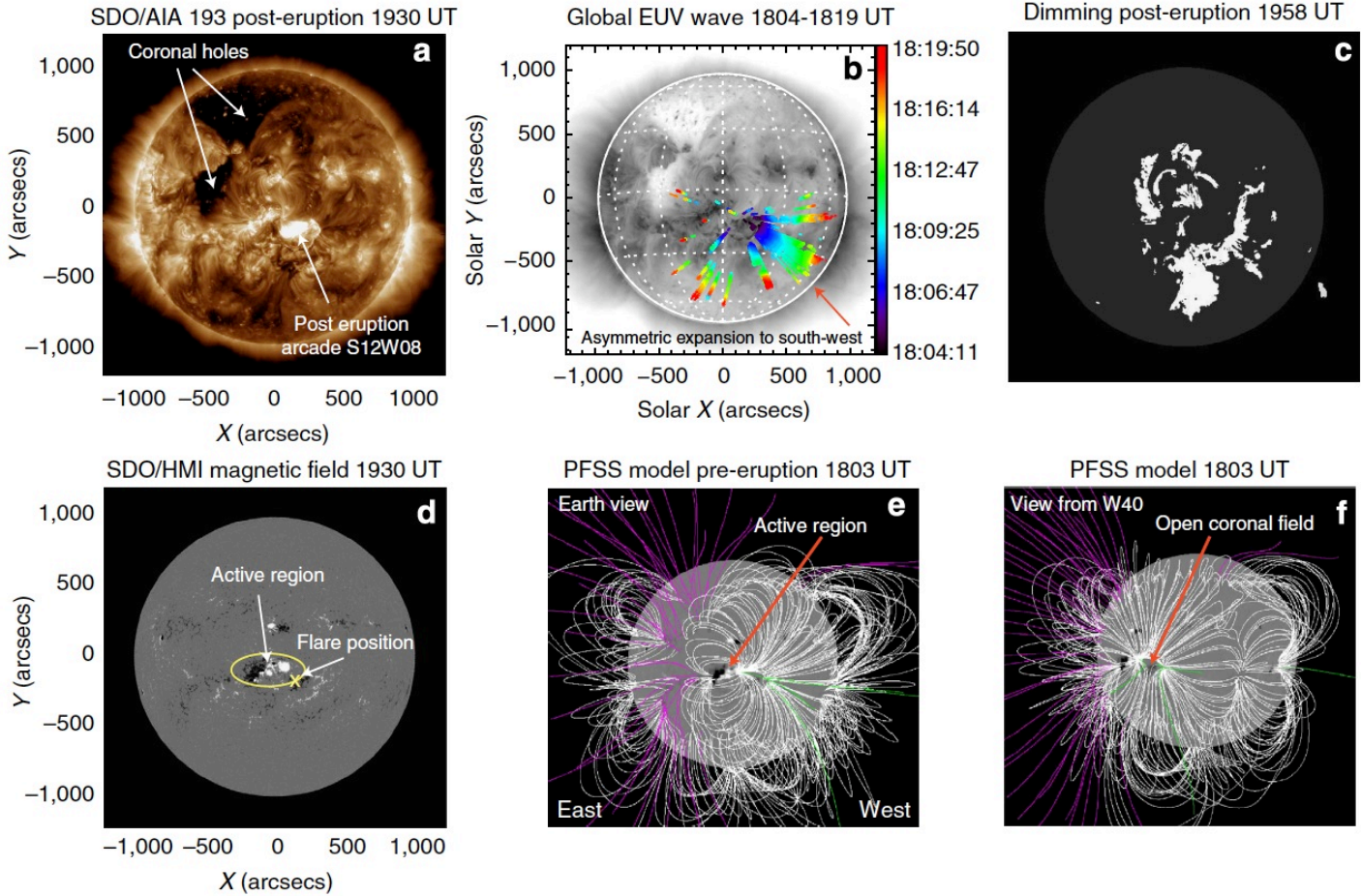

**Figure 1**. **Solar observations of the X1.2 flare and associated phenomena on 7 January 2014 18-20 UT.** (**a**) Location of coronal holes and post eruption arcade in SDO/AIA 193 Å. (**b**) EUV wave evolution between 18:04 and 18:19 UT derived with the CorPITA algorithm. Colors indicate the position of the wave front at different times. (**c**) Final extent of the coronal dimming (SDO/AIA 211 Å). (**d**) SDO/HMI line-of-sight magnetic field, showing the position of the large active region and the flare. White (black) colors indicate positive (negative) magnetic field polarities. (**e, f**) Pre-eruption PFSS model of the solar global magnetic field, as seen from Earth (**e**) and 40° west of Earth (**f**). It depicts closed (white) and open field lines (pink negative polarity, green positive polarity). Solar east (west) is to the left (right) in all images.

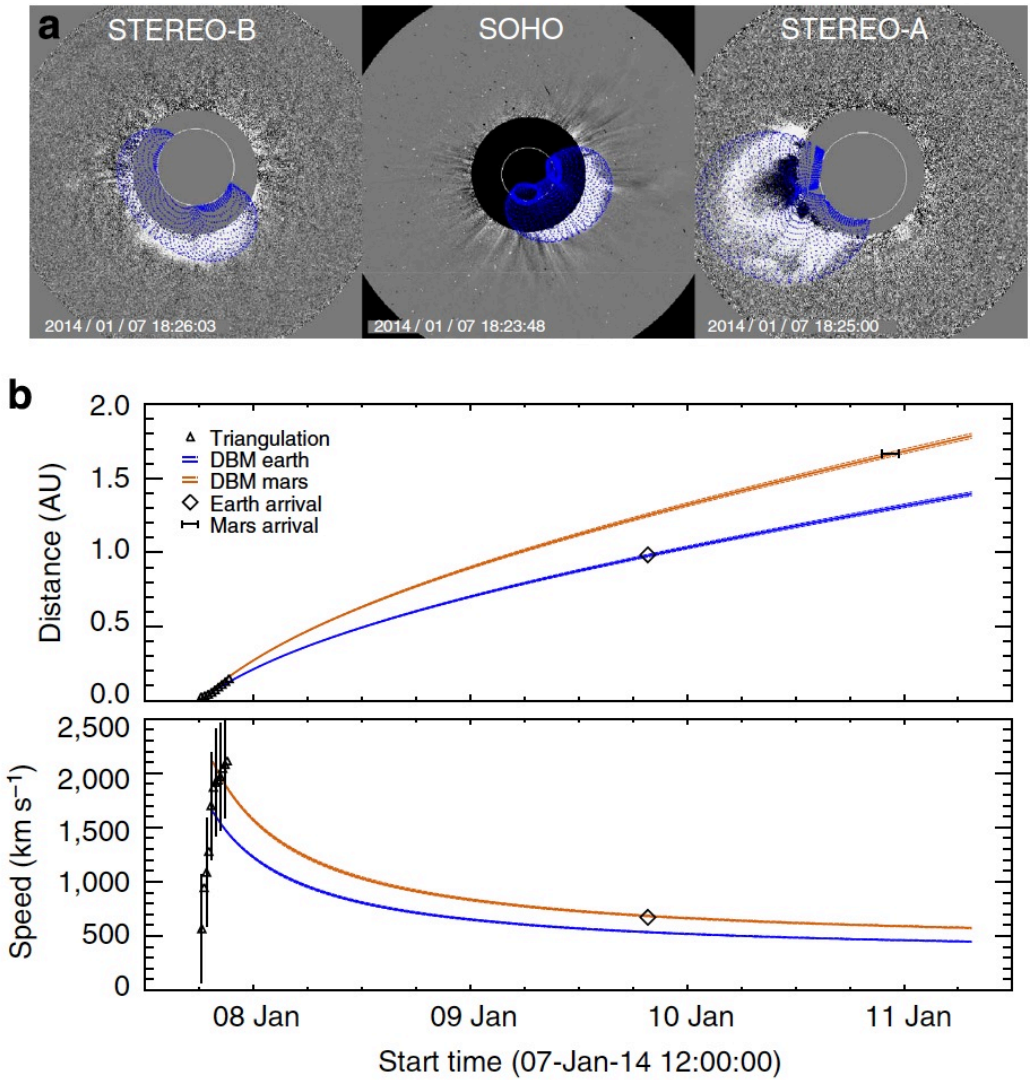

**Figure 2. Graduate Cylindrical Shell model of the CME and interplanetary shock kinematics.** (**a**) Fit of the Graduate Cylindrical Shell (GCS) model (blue grid) overlaid on multipoint coronagraph observations, from left to right: STEREO-B, SOHO, STEREO-A. Shown are results for 7 January 2014 at 18:25 UT ± 1 minute, when the GCS apex was at 4.2 $R_\odot$. (**b**) Distance (top) and speed (bottom) of the CME shock in the ecliptic are shown as a function of time. Blue (orange) solid lines are the kinematics towards Earth (Mars) calculated with the ElEvo model, based on a DBM with $\gamma = 0.165 \times 10^{-7}$ km$^{-1}$ and $w = 400$ km s$^{-1}$. Blue and orange dashed lines indicate errors from a variation of $\gamma$ from 0.16 to 0.17 × 10$^{-7}$ km$^{-1}$, which results from the uncertainty in $t_{Mars}$ of ± 1 hour. Black triangles are the results of triangulation, with speeds and their errors deduced from a derivation of a spline fit on the distance measurements. The observed arrival times and speeds at Earth are indicated by black diamonds, and the arrival window at Mars with a black horizontal line.

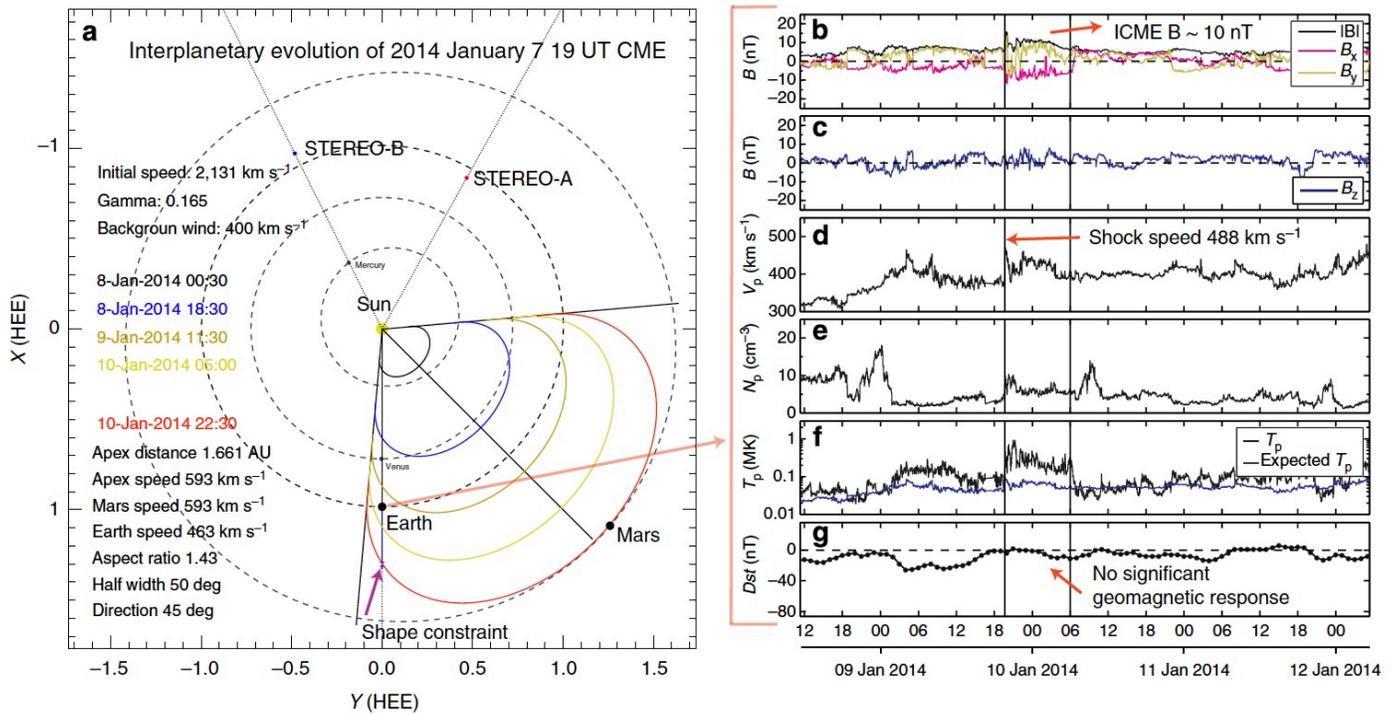

**Figure 3. Ellipse evolution model for the CME shock in the heliosphere and near- Earth solar wind. (a)** Heliospheric positions of various planets and spacecraft on 7 January 2014 at 19:00 UT. The shape of the CME shock given by ElEvo is plotted for different timesteps as indicated by colors. The model parameters (bottom left) are stated for the last timestep $t_{Mars}$ = 10 January 22:30 UT. For the same time, the "shape constraint" gives a window for the heliocentric distance of the ellipse segment along the Sun-Earth line, making the ellipse shape consistent with $t_{Earth}$. **(b-g)** Solar wind magnetic field and bulk proton parameters in near Earth space (Wind SWE/MFI) for 9-11 January 2014. **(b)**: total magnetic field (black) and components $B_x$ (magenta) and $B_y$ (yellow); **(c)** magnetic field $B_z$ component (in Geocentric Solar Ecliptic coordinates); **(d)** proton bulk speed; **(e)** proton density; **(f)** proton temperature (black, expected temperature[40,63] from the solar wind speed in blue); and **(g)** geomagnetic Dst index. The first vertical solid line from the left indicates the arrival of the shock, and the 2nd vertical line delimits the end of the ICME sheath region, which does not seem to be followed by a magnetic ejecta.

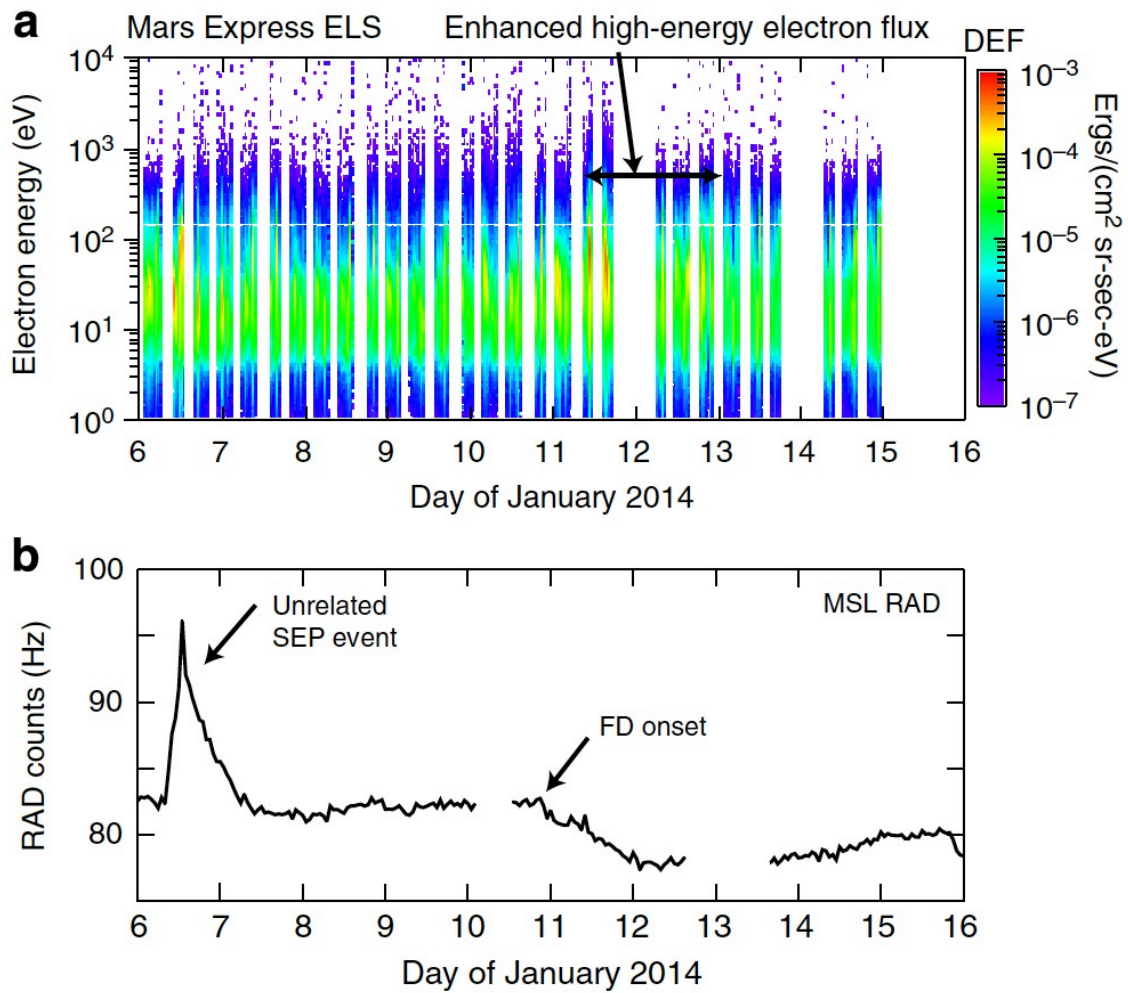

**Figure 4. Observations indicating the CME arrival at Mars.** (**a**) At Mars Express, the CME is observed by the Electron Spectrometer (ELS) as an increase in the electron magnetosheath and solar wind differential energy flux (color coded) starting late on 10 January, with clear enhancements on early 11 January to late 12 January. The horizontal arrow bar delimits the interval of enhanced high energy electrons (30-400 eV) on 11-12 January. (**b**) Counts of energetic particles per second by the RAD experiment on the surface of Mars onboard the Mars Science Laboratory's Curiosity Rover. The high-energy solar energetic particle event stems from an eruption on 6 January. The CME of our study was launched from the Sun on 7 January and its shock hit Mars on 10 January 22:30 UT ± 1 hour, as indicated by the onset of a Forbush decrease of the cosmic ray flux.

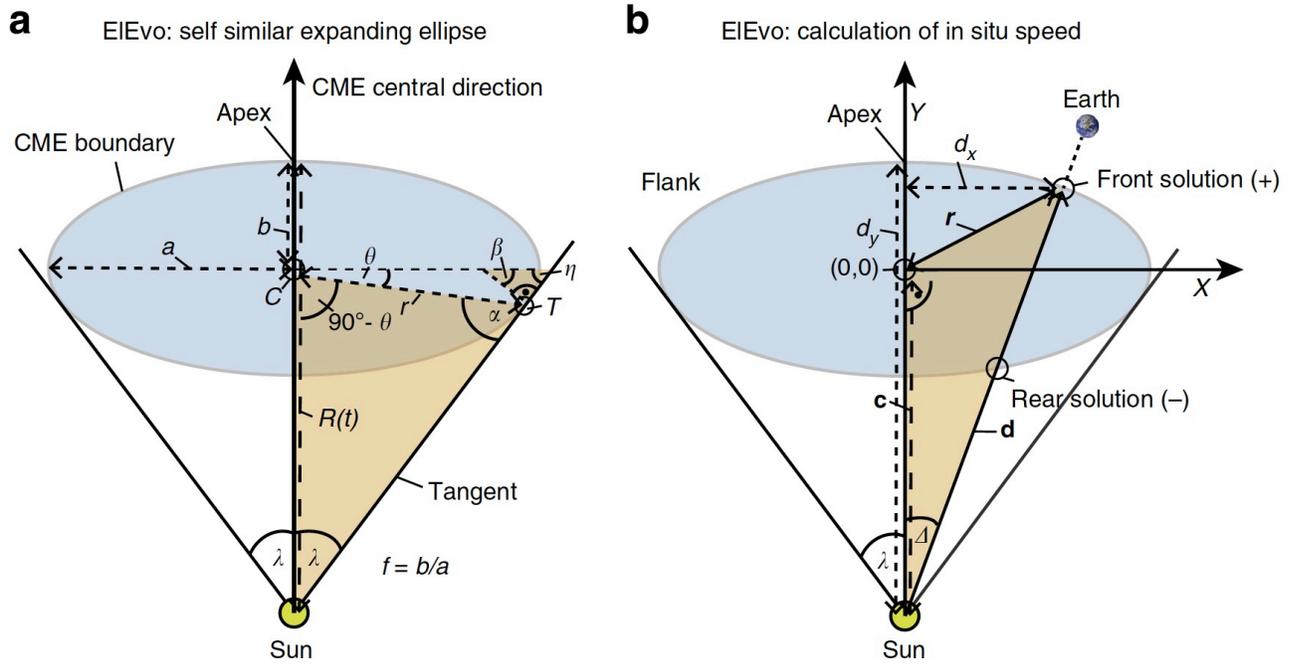

**Figure 5. Derivation of the geometry of a self-similar expanding ellipse in the ElEvo model.** (**a**) A CME leading edge, a shock or the front of a flux rope, described as an ellipse, propagates away from the Sun in the ecliptic or solar equatorial plane with constant angular width and aspect ratio. (**b**) Geometry for deriving the speed of any point along the front of the CME leading edge.

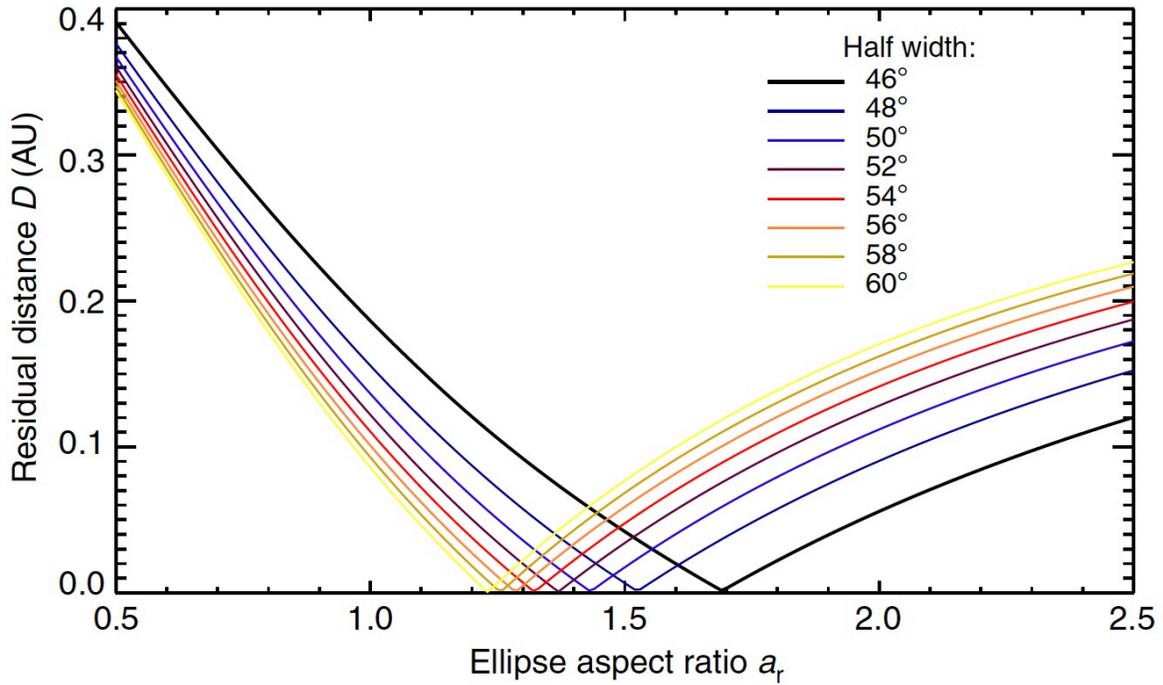

**Figure 6. Optimization of the ElEvo model shape with multipoint in situ observations of a CME shock arrival.** The average residual distances ($D$) between the ellipse and the heliocentric distances of Earth and Mars, at the observed in situ shock arrival times, are plotted as function of the ellipse aspect ratio $a_r$, for half widths of 46 to 60° heliospheric longitude as indicated by the colors. The CME direction is set to W45, the mean value derived from triangulation.